\documentclass{llncs}
\include{macro}
\usepackage{calc,xspace}
\usepackage[modulo]{lineno}  
\usepackage{times}
\usepackage{proof} 
\usepackage{amssymb}
\usepackage{amsmath} 
\usepackage{amsbsy} 
\usepackage{amsfonts} 
\usepackage{euscript} 
\usepackage{graphicx} 
\usepackage{stmaryrd} 
\usepackage{url}

\begin{document}
\pagestyle{plain} 
\author{ Marco Gaboardi}
\institute{Universit\`a di Bologna - INRIA project Focus -
  University of Pennsylvania\\
\email{gaboardi@cs.unibo.it}
}

\title{Linear Dependent Types for Domain Specific Program Analysis\\ (Abstract presented at HOPA 2013)}
\titlerunning{Linear Dependent types for Domain Specific Program
  Analysis\\ (Abstract presented at HOPA 2013)
}
\maketitle      
\begin{abstract}
In this tutorial I will present how a combination of linear and
dependent type can be useful to describe different properties about
higher order programs. Linear types have been proved particularly
useful to express properties of functions; dependent types are useful
to describe the behavior of the program in terms of its control
flow. This two ideas fits together well when one is interested in
analyze properties of functions depending on the control flow of the
program. I will present these ideas with example taken by complexity
analysis and sensitivity analysis. I will conclude the tutorial by
arguing about the generality of this approach.
\end{abstract}

Higher order functional languages provide a powerful abstraction
mechanism helpful both to structure programs in a modular
way and to provide a strong semantics ground to programs. 
However, this same abstraction mechanism makes the analysis of programs more
difficult.  An invaluable tool to ensure properties about  higher order programs are type
system. A typing judgment \verb!|- P : A -> B! gives information
about the function that the program \verb!P! represents, e.g it tells us
that the program \verb!P! represents a function that given as input an
object of type $A$ provide as output an object of type $B$. 

Nowadays, there is a full scale of  type systems that permit to
describe different properties about program. At one end of the scale,
there are \emph{simple types} systems. Simple types, like the ones in the
example above, provide useful information
about the input-output domains of the program and are helpful to ensure
\emph{weak} properties of programs like that programs \emph{cannot go
  wrong}. 
At the other end of the scale, there are systems like intuitionistic
type theories, the type theories behind tools like Coq and Agda, that instead are able to describe very precise
specifications of all the aspects of a program and are helpful to ensure also \emph{strong}
properties of programs. The strength of these systems rely on a
combination of different abstraction mechanisms like dependent
types, polymorphism, inductive and coinductive types and universes. 
The downside of this approach is that the programmer needs to provide an explicit
(partial) proof that his program satisfy the intended property. For
strong properties this task can be overwhelming. 

In between the two ends of the scale there is a full range of other
type systems ideas that have been proved to be useful to analyse
programs. In this tutorial, I want to present some experience in
combining two of these ideas: dependent types and linear types.
From our experience, type systems combining these two ideas 
gives analyses that are useful to prove specific properties of programs.

\medskip

The key idea of dependent type systems is to let the types depend on
the value of terms. 
A particularly simple example of dependent types is the one
represented by \emph{indexed} types~\cite{POPL99*214},~\cite{conf/icfp/ChenX05}. An indexed type
\verb!A[I]! can be seen as the type of elements of type \verb!A! that
meet the property \verb|I|. In general, \verb|I| can be seen as a
boolean predicate even if in some cases it is more convenient to think to
it as a property that identifies one element of \verb|A| (singleton
types). This kind of indexed types are useful to describe more refined
properties of programs. For instance, a typing judgment like 
\begin{verbatim}
i,j : length |- 
      append : listChar[i] -> listChar[j] -> listChar[i+j]
\end{verbatim}
says that  \verb|append| represents a function that takes in
input a first list of character of length \verb|i|, a second list of character of length \verb|j| and returns a list of
character of length \verb|i+j|. 
Indexed type systems are very useful to describe the input-output behavior of
programs, however when one is interested in describing the properties
of functions this approach has some drawbacks.  

As an example of this
fact, let's consider \emph{function sensitivity}.
The sensitivity of a function $f(x)$ is an upper bound on how much $f(x)$
can change in response to a change to $x$---in other words, if $f$ has sensitivity
$k$, then $|f(x+\delta)-f(x)|\le k\cdot |\delta|$ for all $x$ and $\delta$. This
property,  also 
known as \emph{Lipschitz continuity}, can be extended to 
entire programs with multiple inputs, and it has important applications in many
parts of computer science, including control theory, dynamic systems, 
program analysis, and data privacy.

Indexed types, can be used to ensure function sensitivity, e.g. a
typing judgment as 
\begin{center}
  \verb!i : real |- P : R[i] -> R[2*i+1]!
\end{center}
can ensure that the function computed by \verb|P| is
$2$-sensitive. The way we can deduce this is by looking at the fact
that the function $f$ computed by \verb|P| is such that $f(x)= 2*x+1$
and by knowing that $f$ is $2$ sensitive. This kind of reasoning has
some drawbacks: first, the reasoning on the sensitivity is external to
the type system. i.e. this is performed on the semantics of the
program; second, to capture the precise input-output behavior, 
the types have to be very rich---this makes difficult to
perform the analysis in an automatic way; 
third, to generalize the analysis to study the sensitivity on function
spaces---useful for programs like \verb|map|, \verb|fold|, etc.---we
need to extend indexed types to functions.
One possible solution is to consider full dependent type systems, like
intuitionistic type theory. However, this approach once again makes
difficult to perform the analysis in an automatic way.

An  alternative is to use other type ideas specifically developed to
talk about properties of functions. One important example is
represented by \emph{linear types}. One of the main idea of linear
logic---from which linear types systems are inspired--- is to
decompose a function type as \verb| A -> B = !A -o B|. A standard
description of this decomposition says that \verb|A -o B| can be seen as a
the type of a function that uses its argument exactly once, while
\verb|!A -o B| can be seen as the type of a function that uses its
argument an arbitrary number of times. A less standard description
(but immediate from the semantics of linear logic)
says that the \emph{linear} function space constructor \verb|-o| is a mapping from
an \emph{individual} of the input to an \emph{individual} of the output;
while  the \emph{modality} \verb|!| describes how the individuals of the type \verb|A| are combined to obtain the
individuals of the type \verb|!A|.
The important point here is that now the modality can be used to
capture  the property of the function space. To provide a finer
analysis of this function space, we
generalize the modality $!$ to a set of indexed-modalities $!_I$. 

Let's see how we can use this idea to describe function
sensitivity. First, we can say that the linear map like \verb|A -o B| represents
a $1$-sensitive function from \verb|A| to \verb|B|. Second, by taking
an index $I$ to be equal to a number $r$ we can say
that the modality \verb|!_r A| is the same type as \verb|A| except
that the distance between elements in \verb|A| is now scaled by
$r$. By building the type system accordingly to this idea, we can
ensure that a term to which we can assign type \verb|!_r A -o B|
represents an $r$-sensitive function from \verb|A| to \verb|B|.
As an example, let's assume that \verb|+| is a $1$-sensitive function
in its arguments, i.e. %
\verb|+ : !_1 R -o !_1 R -o R| then we have a judgment%
\begin{center}
  \verb!|- !$\lambda$\verb|x. x + x :!_2 R -o R|
\end{center}
ensuring that the function $\lambda$\verb|x. x + x| is $2$-sensitive
in its argument. This example also shows why this property can be
described by linear types. Indeed, the sensitivity of a function depends on the
single uses of the input.

A type system built around this idea have been used in~\cite{fuzz} as
the basic block of a language for differential privacy. This type
system provides an efficient analysis tool to infer the sensitivity of
higher order programs~\cite{fuzzinf}. Unfortunately, this analysis
fails on many important programs. The main problem is that the
sensitivity of the program on one input can depends on some other
input. Consider as an example an \verb|iter| function that given a
function $f$ maps it on an input value $k$, a number
of time specified by a parameter $n$, i.e. \verb!iter!$\ n\ f\
k=f^n(k)$.
 We can assume to have the
following type:
 \begin{verbatim}
           |- iter : Nat -> (R -> R) -> R -> R
\end{verbatim}
The problem here is that the sensitivity of \verb!iter! on $k$ depends on the value of $n$ and on the
sensitivity of $f$, and this kind of dependency cannot be captured by the
indexed-modality $!_r$ alone.

The natural way to generalize the analysis proposed in \cite{fuzz} is to combine it with
indexed types and have a similar set of indexes both in types
and in modalities. For instance, we can have a judgment:
\begin{verbatim}
|- iter : Nat[i] -> (!_r R -o R) -> !_(r*i) R -o R
\end{verbatim}
saying that \verb|iter| is \verb!r*i! sensitive on its argument, where
\verb|i| is the size of $n$ and $r$ is the sensitivity of $f$ and
where we use the standard type \verb|->| to omit the sensitivities of
the other arguments.
A type system able to perform this kind of analysis has been used
in~\cite{Dfuzz} to extend the language for differential privacy
described in~\cite{fuzz}.
The resulting system besides combining the indexed type approach with
the one of linear types has also some other features like subtyping
and quantifiers. The gain in using this generalized approach is the ability to analyse
more general programs where the property of interest depends on the
control flow of the program.

An important motivation for following the approach described above is given by type
inference and type checking. Type checking and inference for linear
dependent types can be seen as an extension of the usual ML type
checking and inference by two extra phase where the constraint
generation and the constraint resolution on the index language are
performed. This suggest clearly, a two step approach where constraints
are first generated by a standard algorithm and then are passed to an
automatic solver. In this way, the strength of the analysis is reduced
to the strength of the solver in deciding the generated constraints.

The idea of combining linear and dependent types, dubbed naturally 
\emph{linear dependent} types, has been originally proposed by myself and Dal Lago to
perform complexity analysis of higher order
programs~\cite{DLG10}\cite{DLG12}.
Indeed, an analysis for the complexity of programs need to be able to
express it as a function of the input values. Moreover, every precise
description of the complexity has to consider the control flow of the
program.
These two considerations motivated the combination of linear types,
already extensively used in the area of Implicit Computational
Complexity, with indexed types. The kind of types used in~\cite{DLG10}
are more general than the one I outlined above. Indeed, in
\cite{DLG10} a modal type has the shape $!_{a<I}A$. So, it 
is indexed not only by an index $I$ (representing a natural number) but instead by an inequality
$a<I$ that says that the variable $a$, that can appear in $A$ can
assume values less that $I$. 
More precisely, for complexity we can think to the type  $!_{a<I}A$ as representing
the type $A[0/a]\otimes A[1/a]\otimes\cdots\otimes A[I-1/a]$. This
generalized form of modality permits to have the value of some modality
to depend also on values of other modalities. The system presented
in~\cite{DLG10} has also other components that make it
non-standard. In particular, the system is parametrized on an
equational program providing the semantics of the indexes. All those
components are needed to obtain a \emph{relative completeness} result stating that
the complexity analysis is complete, assuming an oracle to decide the constraints.

The complexity analysis described in~\cite{DLG10} is in term of a call
by name Krivine-like abstract machine.  Dal Lago and
Petit~\cite{DLP12a} have shown how the same approach can be used to
give precise information also for a call by value CEK-like abstract
machine. Moreover, in~\cite{DLP13} they also have shown how the
relative completeness result of the type system in~\cite{DLG10} can be
transferred to a relative completeness result of the type inference
process.

The pattern described above is not limited to \emph{sensitivity} and
\emph{complexity}. At the end of the tutorial I will show how this
idea can be applied also to design type systems permitting other
analysis like information flow. The important foundational aspect of
this approach is that the operation involved in the constraints on
the indexes correspond to the operations needed to reason about the
property of interest, and that these operations in turn correspond  to different interpretations of the laws of
Linear Logic (contraction, weakening, digging and dereliction).

\bibliographystyle{plain}

\end{document}